\documentclass[conference, a4paper]{IEEEtran}
\usepackage{cite}
\usepackage[latin1]{inputenc}
\usepackage[english]{babel}
\usepackage{amsmath,amsfonts,amssymb}
\usepackage{graphicx, epsfig}
\usepackage{subfigure}
\usepackage{array}
\usepackage{multirow}
\usepackage{mdwmath}
\usepackage{mdwtab}
\usepackage{algorithmic}
\usepackage{algorithm}

\usepackage{color}

\ifCLASSINFOpdf

\else

\fi

\begin{document}
%
\title{\huge Repeat Accumulate Based Designs for LDPC Codes on Fading Channels}

\author{\IEEEauthorblockN{Andre G. D. Uchoa, Cornelius Healy and Rodrigo C. de Lamare}
\IEEEauthorblockA{University of York,\\
Heslington York, YO10 5DD, UK\\
$\lbrace$agdu500, cth503, rcdl500$\rbrace$@york.ac.uk}}

\maketitle

\begin{abstract}
Irregular repeat-accumulate Root-Check LDPC codes based on
Progressive Edge Growth (PEG) techniques for block-fading channels
are proposed. The proposed Root-Check LDPC codes are
 {both suitable for channels under $F = 2, 3$
independent fadings per codeword and} for fast fading channels. An
IRA(A) Root-Check structure is devised for $F = 2, 3$ independent
fadings. The performance of the new codes is investigated in terms
of the Frame Error Rate (FER). Numerical results show that the IRAA
LDPC codes constructed by the proposed algorithm
 {outperform by about 1dB the existing} IRA Root-Check
LDPC codes under fast-fading channels.
\end{abstract}


%
\IEEEpeerreviewmaketitle

\section{Introduction}
Due to multi-path propagation and mobility, wireless systems are
characterized by time-varying channels with fluctuating signal
strength. In applications subject to delay constraints and
slowly-varying channels, only limited independent fading
realizations are experienced. In such conditions also known as
non-ergodic scenarios, the channel capacity is zero since there is
an irreducible probability, termed outage probability
\cite{rasmussen.10}, that the transmitted data rate is not supported
by the channel. The case of interest in this work is the
block-fading type \cite{rappaport}. A simple and useful model that
captures the essential characteristics of non-ergodic channels is
the block-fading channel \cite{rappaport}. It is especially
important in wireless communications with slow time-frequency
hopping (e.g., cellular networks and wireless local area networks)
or multi-carrier modulation using Orthogonal Frequency Division
Multiplexing (OFDM) \cite{boutros.07}. Codes designed for
block-fading channels are expected to achieve the limited channel
diversity and to offer good coding gains.

In \cite{boutros.07} the authors proposed a family of LDPC codes
called Root-Check for block-fading channels. Root-check codes are
able to achieve the maximum diversity of a block-fading channel and
have a performance near the limit of outage when decoded using the
Sum Product Algorithm (SPA). Root-check codes are always designed
with code rate $R = 1/F$, since the Singleton bound determines that
this is the highest code rate possible to obtain the maximum
diversity order \cite{boutros.07}.
Li and Salehi \cite{salehi.10} proposed the construction of
structured Root-Check LDPC codes via circulating matrices, i.e.,
Quasi-Cyclic LPDC codes (QC-LDPC). In \cite{salehi.10} the authors
have shown that the QC-LDPC codes are as good as the randomly
generated Root-Check LDPC codes on block-fading channels. It is
known that the girth, the length of the shortest cycle in the graph
of this code has a significant effect on the performance of the
code. Among the algorithms capable of producing high performance
LDPC codes for short to medium lengths is the Progressive Edge
Growth (PEG) algorithm \cite{hu.05}.



In order to improve the girth of the Root-Check LDPC codes the
PEG-Based Root-Check LDPC codes \cite{iswcs.11, peg.comms.11,
iswcs.12} which are designed in a PEG based technique \cite{dopeg}
have been developed. The proposed PEG-Based Root-Check LDPC codes
presented in \cite{iswcs.11, peg.comms.11, iswcs.12} outperformed
other LDPC codes based on the Root-Check structure. The best result
presented by the works in \cite{iswcs.11, peg.comms.11, iswcs.12}
reinforces that a Root-Check LDPC code generated with an algorithm
based on PEG produces a better performance than that of a
 {standard design.}


The accumulator-based codes that were invented first are the
so-called repeat-accumulate (RA) codes \cite{ryanbook}. Despite
their simple structure, they were shown to provide good performance
and, more importantly, they paved a path toward the design of
efficiently encodable LDPC codes. The contribution of this paper is
to present a PEG-based algorithm to design IRA Based LDPC codes with
Root-Check properties for block-fading channel with $F = 2, 3$
fadings per coded word. The key points in using IRA Based Root-Check
LDPC codes are the simplicity in designing such codes, faster
encoding than  {conventional LDPC} methods and flexibility in terms
of  {rate compatibility as provided} by intentional puncturing
strategies \cite{ryanbook}. A strategy that imposes irregular
repeat-accumulate and Root-Check constraints on a PEG-based
algorithm is devised. The codes generated by the proposed algorithm
can achieve a very good performance in terms of Frame Error Rate
(FER) with respect to the theoretical limit. The
 {proposed} design can save up to 1dB in terms of
signal to noise ratio (SNR) to achieve the same FER when compared to
other codes under fast fading channels. For the case of block-fading
channels, the codes achieve  {a comparable} performance in terms of
FER as the works in \cite{iswcs.11, peg.comms.11, iswcs.12}.


The rest of this paper is organized as follows. In Section II we define the system model, while in Section III we present the structure of a RA-Root-Check LDPC codes for $F = 2, 3$ fadings. In Section IV we introduce the proposed PEG-based algorithm. Section V presents numerical results, while Section V concludes the paper.

\section{System Model}

Consider a block fading channel, where $F$ is the number of
independent fading blocks per codeword of length $N$. Following
\cite{salehi.10}, the {\it t}-th received symbol is given by:
\begin{equation} \label{eq:recsymbol}
 r_{t} = h_{f}s_{t}+n_{g_{t}},
\end{equation}
where $t=\{1,2,\cdots,N\}$, $f=\{1,2,\cdots,F\}$, $f$ and $t$ are
related by $f=\lceil F\frac{t}{N}\rceil$, where $\lceil \phi \rceil$
returns the smallest integer not smaller than $\phi$, $h_f$ is the
real Rayleigh fading coefficient of the $f$-th block, $s_{t}$ is the
transmitted signal, and $n_{g_{t}}$ is additive white Gaussian noise
with zero mean and variance $N_0/2$. In the case of fast fading we
assume that each received symbol $r_{t}$ will be under a distinct
fading coefficient, which means $F = N$. In this paper, we assume
that the transmitted symbols $s_{t}$ are binary phase shift keying
(BPSK) modulated. We assume that the receiver has perfect channel
state information, and that the SNR is defined as $E_b/N_0$, where
$E_b$ is the energy per information bit. The information
transmission rate is $R=K/N$, where $K$ is the number of information
bits per codeword of length $N$. For the case of a block-fading
channel, we consider $R=1/F$, since then it is possible to design a
practical diversity achieving code \cite{salehi.10}. The performance
of a communication system in a non-ergodic block-fading channel can
be investigated by means of the outage probability \cite{salehi.10},
which is defined as:
\begin{equation} \label{eq:pout}
P_{out} = {\cal P}( I < R ),
\end{equation}where ${\cal P}(\phi)$ is the probability of event $\phi$. The mutual information $I_{G}$, for Gaussian channel inputs is\cite{salehi.10}:
\begin{equation}\label{eq:outage}
I_{G}=\frac{1}{F} \sum_{f=1}^F \frac{1}{2}\log_2 \left( 1 + 2R\frac{E_b}{N_0}h_f^2\right),
\end{equation}so that an outage occurs when the average accumulated mutual information among blocks is smaller than the attempted information transmission rate.

\section{RA Based LDPC Codes Design}

\begin{figure}[htb]
 \centering
\resizebox{88mm}{!}{
\includegraphics{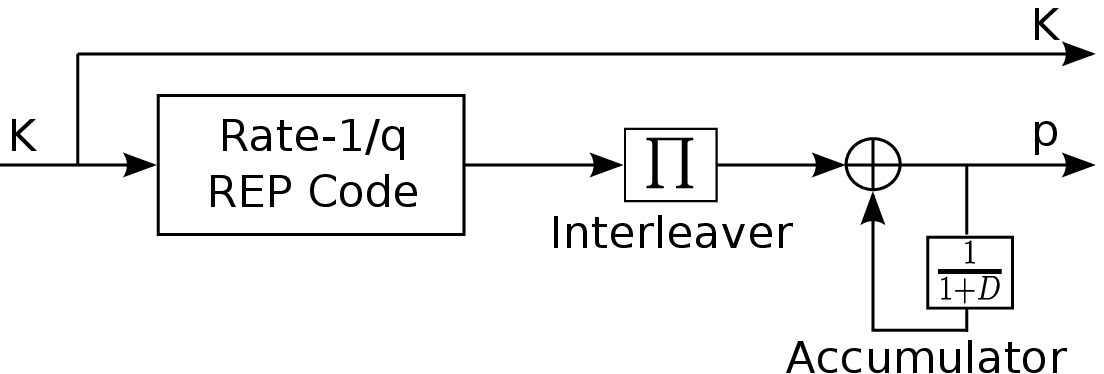}
}
\caption{A systematic repeat-accumulate code block diagram, where $K$ is the number of information bits and $p$ denotes the parity bits.}
\label{fig:ra_diagram}
\end{figure}

A repeat-accumulate (RA) code consists of a serial concatenation,
through an interleaver, of a single rate $1/q$ repetition code with
an accumulator having transfer function $\frac{1}{1+D}$, where $q$
 { is the number of repetitions for each group of $K$
information bits}. Fig. \ref{fig:ra_diagram} shows a typical
repeat-accumulate code block diagram. The implementation of the
transfer function $\frac{1}{1+D}$ is identical to an accumulator,
although the accumulator value can be only $0$ or $1$ since the
operations are over  {the binary field} \cite{ryanbook}. As
discussed in \cite{ryanbook}, to ensure a large minimum Hamming
distance, the interleaver should be designed so that consecutive 1s
at its input are widely separated at its output. The RA based codes
discussed throughout this paper will be systematic. The main
limitation of RA codes are the code rate, which cannot be higher
than $1/2$. Also, these codes are not capacity-approaching.
Therefore, we will be discussing in the next subsections some
enhancements to RA codes which permit operation closer to the
block-fading outage probability.

\subsection{IRA Root-Check Design}

Irregular repeat-accumulate (IRA) codes generalize the concept of RA
codes by changing the repetition rate  {for each group of $K$
information bits and performing a linear combination of the repeated
bits which are sent through the accumulator}. Furthermore, IRA codes
are typically systematic. IRA codes allow flexibility in the choice
of the repetition rate for each information bit so that high-rate
codes may be designed. And, their irregularity allows operation
closer to the capacity limit \cite{ryanbook}.

The Parity Check Matrix (PCM) for systematic RA and IRA codes has
the form $\mathbf{H} = \left[\mathbf{H}_u~\mathbf{H}_p\right]$,
where $\mathbf{H}_p$ is a square dual-diagonal matrix given by
\begin{equation}
 \mathbf{H}_p =
    \begin{bmatrix}
    1 &  &  & & \\
    1 & 1 & & & \\
      & \ddots & \ddots & &\\
      & & 1 & 1 & \\
      & &  & 1 & 1\\
    \end{bmatrix}.
\label{eq:pcm_ra_ira}
\end{equation}
For RA codes, $\mathbf{H}_u$ is a regular matrix having column
weight $q$ and row weight $1$. For IRA codes, $\mathbf{H}_u$ has
irregular columns and rows weights. The Generator Matrix (GM) can be
obtained as $\mathbf{G} =
\left[\mathbf{I}~\mathbf{H}_{u}^{T}\mathbf{H}_{p}^{-T}\right]$,
where $\mathbf{I}$ is an identity matrix of size $K \times K$. In
matrix notation $\mathbf{H}_{p}^{-T}$ can be
 {described as}
\begin{equation}
 \mathbf{H}_{p}^{-T} =
    \begin{bmatrix}
    1 & 1 & \cdots & & 1\\
     & 1 & 1 & \cdots & 1\\
      & & \ddots & & \vdots\\
      & & & 1 & 1\\
      & &  & & 1\\
    \end{bmatrix}.
\label{eq:hpinvt}
\end{equation}

\subsubsection{IRA Root-Check Rate $\frac{1}{2}$}

To design a Root-Check with an IRA structure we have imposed some
constraints in terms of parity check matrix to guarantee the
Root-Check properties. Following the notation adopted in
\cite{iswcs.11}, for the case of a systematic Rate $1/2$ with $F =
2$, the PCM must be like
\begin{equation}
 \mathbf{H}=
    \begin{bmatrix}
    \mathbf{I} & \mathbf{H}_{2} & \mathbf{0} & \mathbf{H}_{3}\\
    \mathbf{H}_{2} & \mathbf{I} & \mathbf{H}_{3} & \mathbf{0}\\
    \end{bmatrix},
\label{eq:pcm_rc_r12}
\end{equation}
where $\mathbf{H}_{2}$ and $\mathbf{H}_{3}$ are $\frac{N}{2}\times\frac{N}{2}$ sub-matrices with Hamming weight two and three respectively, while $\mathbf{0}$ is a null sub-matrix. Therefore, to impose the RA structure and Root-Check properties the PCM of an IRA Root-Check is
\begin{equation}
 \mathbf{H}=
    \begin{bmatrix}
    \mathbf{I} & \mathbf{H}_{2} & \mathbf{0} & \mathbf{H}_{p}\\
    \mathbf{H}_{2} & \mathbf{I} & \mathbf{H}_{p} & \mathbf{0}\\
    \end{bmatrix},
\label{eq:pcm_ira_rc_r12}
\end{equation}
where $\mathbf{H}_{p}$ is a dual diagonal matrix with size
$\frac{N}{2}\times\frac{N}{2}$. By doing this, we are able to
achieve the diversity of the channel and the same performance in
terms of FER as the codes designed in \cite{iswcs.11, peg.comms.11}.

\subsubsection{IRA Root-Check Rate $\frac{1}{3}$}

 {For the} case of Rate $1/3$ with $F = 3$, we follow
a similar structure to the one adopted in \cite{iswcs.12,
boutros.07}. However, we have made some modifications on the
accumulator to  {approach the outage probability}. The accumulator
used in this case has a transfer function
$\frac{1}{1+D+D^{\frac{N}{9}}}$ as suggested in
\cite{sarah.johnson.05}. As a result, $\mathbf{H}_{p}$ must be
redefined as
\begin{equation}
 \mathbf{H}_{p} =
    \begin{bmatrix}
    \mathbf{H}_{p1} \\
    \mathbf{H}_{p2} \\
    \end{bmatrix},
\label{eq:hpira3}
\end{equation}

\begin{equation}
\mathbf{H}_{p1} =
 \begin{bmatrix}
 1 & 0 & \cdots & \cdots & \cdots & \cdots & \cdots & 0\\
 1 & 1 & 0 & \ddots & \ddots & \ddots & \ddots & 0\\
 \vdots & \ddots & \ddots & 0 & \ddots & \ddots & \ddots & 0\\
  0 & 0 & 1 & 1 & 0 & 0 & 0 & 0\\
 \end{bmatrix},
\label{eq:hp1}
\end{equation}

\begin{equation}
\mathbf{H}_{p2} =
 \begin{bmatrix}
  1 & 0 & \cdots & 0 & 1 & 0 & \cdots & 0\\
  0 & \ddots & 0 & 0 & 1 & 1 & 0 & 0\\
  \vdots & 0 & \ddots & 0 & 0 & \ddots & \ddots & 0\\
  0 & 0 & 0 & 1 & 0 & 0 & 1 & 1\\
 \end{bmatrix},
 \label{eq:hp2}
\end{equation}
where $\mathbf{H}_{p1}$ and $\mathbf{H}_{p2}$ are sub-matrices with
dimensions $\frac{N}{9}\times\frac{2N}{9}$. Now, we can show the
final PCM $H = [\mathbf{H}_{u} \vert \mathbf{H}_{p}]$ for the case
of an IRA Root-Check Rate $1/3$ as
\begin{equation}
 \mathbf{H}=
    \begin{bmatrix}
     \mathbf{I} & \mathbf{H}_{1} & \mathbf{0} & \vert & \mathbf{0} & \mathbf{H}_{p2} & \mathbf{0}\\
     \mathbf{I} & \mathbf{0} & \mathbf{H}_{1} & \vert & \mathbf{0} & \mathbf{0} & \mathbf{H}_{p1}\\
     \mathbf{H}_{1} & \mathbf{I} & \mathbf{0} & \vert & \mathbf{H}_{p1} & \mathbf{0} & \mathbf{0}\\
     \mathbf{0} & \mathbf{I} & \mathbf{H}_{1} & \vert & \mathbf{0} & \mathbf{0} & \mathbf{H}_{p2}\\
     \mathbf{H}_{1} & \mathbf{0} & \mathbf{I} & \vert & \mathbf{H}_{p2} & \mathbf{0} & \mathbf{0}\\
     \mathbf{0} & \mathbf{H}_{1} & \mathbf{I} & \vert & \mathbf{0} & \mathbf{H}_{p1} & \mathbf{0}\\
    \end{bmatrix},
\label{eq:pcm_ira_rc_r13}
\end{equation}
where $\mathbf{H}_{1}$ and $\mathbf{I}$ are sub-matrices with
 { dimensions $\frac{N}{9}\times\frac{N}{9}$ and
$\mathbf{H}_{1}$ is a sub-matrix with Hamming weight equal to $1$.
The null sub-matrices $\mathbf{0}$ in the right hand side of
(\ref{eq:pcm_ira_rc_r13}) have dimensions
$\frac{N}{9}\times\frac{2N}{9}$ while in the left hand side the
dimensions are $\frac{N}{9}\times\frac{N}{9}$.  The reason why we
split $\mathbf{H}_p$ into two sub-matrices $\mathbf{H}_{p1}$ and
$\mathbf{H}_{p2}$ is to guarantee the full rank and full diversity
properties stated in \cite{boutros.07}.}

\subsection{IRAA Root-Check Design}

\begin{figure}[htb]
 \centering
\resizebox{88mm}{!}{
\includegraphics{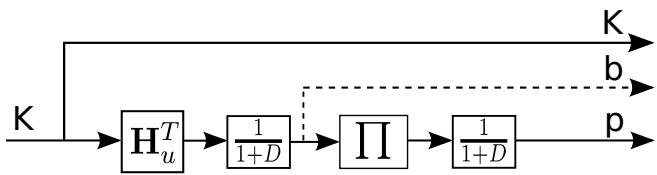}
}
\caption{An systematic irregular repeat-accumulate and accumulate code block diagram. Where $K$ are the information bits, $b$ and $p$ are the parity bits.}
\label{fig:iraa_diagram}
\end{figure}

The general structure of an Irregular Repeat-Accumulate and
Accumulate (IRAA) encoder can be seen in Fig.
\ref{fig:iraa_diagram}. In this figure, we see an extra parity bits
which are termed $b$ and the normal parity bits $p$. The $b$ parity
bits can be punctured to obtain a higher code rate. For instance, in
general an IRAA code is rate $1/3$ without puncturing, while
puncturing $b$ it can be obtained a code with rate $1/2$.

The PCM of an IRAA LDPC code can be represented by
\begin{equation}
\mathbf{H} =
\begin{bmatrix}
    \mathbf{H}_{u} & \mathbf{H}_{p} & \mathbf{0}\\
    \mathbf{0} & \prod_{1} & \mathbf{H}_{p}\\
\end{bmatrix},
\label{eq:pcm_iraa}
\end{equation} where $\prod_{1}$ must be a sub-matrix with rows and columns with Hamming weight one.

 {In order to obtain IRAA Root-Check LDPC codes some
constraints must be imposed on the standard IRAA design. We have
noticed that the IRAA Root-Check LDPC codes led to a more flexible
rate compatible code and a better performance under fast fading
channels. This will be illustrated by simulations later on in
Section V.}

\subsubsection{IRAA Root-Check Rate $\frac{1}{2}$}
We applied the Root-Check structure from (\ref{eq:pcm_ira_rc_r12}) in (\ref{eq:pcm_iraa}) to obtain the following PCM for rate $1/2$
\begin{equation}
\mathbf{H} =
\left[
\begin{array}{cccccc}
\mathbf{I} & \mathbf{H}_{2} & \mathbf{0} & \mathbf{H}_{p} & \mathbf{0} & \mathbf{0}\\
\mathbf{H}_{2} & \mathbf{I} & \mathbf{H}_{p} & \mathbf{0} & \mathbf{0} & \mathbf{0}\\
\mathbf{0} & \mathbf{0} & \multicolumn{2}{c}{\multirow{2}{*}{$\prod_{1}$}} & \mathbf{0} & \mathbf{H}_{p}\\
\mathbf{0} & \mathbf{0} & & & \mathbf{H}_{p} & \mathbf{0}\\
\end{array}
\right],
\label{eq:pcm_iraa_rc_r12}
\end{equation}
where $\mathbf{I}$, $\mathbf{H}_{2}$, $\mathbf{H}_{2}$ and
$\mathbf{0}$ are all $\frac{N}{9}\times \frac{N}{9}$ in dimension,
while $\prod_{1}$ is $\frac{N}{3}\times \frac{N}{3}$. The key point
to guarantee the full diversity property is the puncturing
procedure. Instead of puncturing $b$ parity bits we have punctured
$p$. The reason why puncturing $p$ instead of $b$ guarantees the
full diversity is due to the fact that the Root-Check structure of
the code is  {kept unchanged}. For the case of fast fading we have
punctured in the same manner as for the case of block-fading
channels.

\subsubsection{IRAA Root-Check Rate $\frac{1}{3}$}

For the case of rate $1/3$ we considered the design done in
(\ref{eq:pcm_ira_rc_r13}) and we apply the constraints in
(\ref{eq:pcm_iraa}) to obtain the following PCM
\begin{equation}
\mathbf{H} =
\begin{bmatrix}
    \mathbf{H}_{u} & \vert & \mathbf{H}_{p} & \mathbf{0}\\
    \mathbf{0} & \vert & \prod_{1} & \mathbf{H}_{p}\\
\end{bmatrix}.
\label{eq:pcm_iraa_rc_r13}
\end{equation}
It must be noted that without puncturing the code rate is $1/5$.

\section{Proposed Design Algorithm}

Here, we introduce some definitions and  {a specific notation}.
Then, we present the pseudo-code of our proposed algorithm. In this
work, the  {scenarios of a block-fading channel with $F = 2$, $F =
3$ and a fast fading channel} are considered. In extending to a
greater number of fadings, $F>4$, the general structure presented is
maintained. The LDPC code in systematic form is specified by its
sparse PCM $\mathbf{H} = [\mathbf{A}\vert \mathbf{B}]$, where
$\mathbf{A}$ is a matrix of size $M \times K$, and $\mathbf{B}$ is
an $M \times M$ matrix.  {The generator matrix (GM)} for the code is
$\mathbf{G} = [\mathbf{I}\vert(\mathbf{B}^{-1}\mathbf{A})^{T}]$,
$\mathbf{I}$ is an identity matrix of size $K\times K$.

The variable node degree sequence $D_s$ is defined as the set of
column weights of the designed \textbf{H}, and is prescribed by the
variable node degree distribution $\lambda(x)$ as described in
\cite{richardson.01}. Moreover, $D_s$ is arranged in non-decreasing
order. The proposed algorithm, called IRA-PEG Root-Check, constructs
\textbf{H} by operating progressively on variable nodes to place the
edges required by $D_s$. The Variable Node (VN) of interest is
labelled $v_j$ and the candidate check nodes are individually
referred to as $c_i$. The IRA-PEG Root-Check algorithm chooses a
check node $c_i$ to connect to the variable node of interest $v_j$
by expanding a constrained sub-graph from $v_j$ up to maximum depth
$l$. The set of check nodes found in this sub-graph is denoted
$N_{v_j}^l$ while the set of check nodes of interest, those not
currently found in the sub-graph, are denoted
$\overline{N_{v_j}^l}$.

\subsection{Pseudo-code for the IRA-PEG Root-Check Algorithm}
Initialization: A matrix of size $M \times N$ is created with the identity matrices $\mathbf{I}$ and parity matrices $\mathbf{H}_{p}$ in the positions shown in (\ref{eq:pcm_ira_rc_r12}), (\ref{eq:pcm_ira_rc_r13}), (\ref{eq:pcm_iraa_rc_r12}), (\ref{eq:pcm_iraa_rc_r13}) and zeros in all other positions. We define the indicator vectors ${\mathbf{z}}_{1}, \cdots, {\mathbf{z}}_{F}$ for the cases $R = \frac{1}{2}$, $R = \frac{1}{3}$ respectively as:
\begin{eqnarray}
{\mathbf{z}}_{1} & = & [\mathbf{0}_{1 \times \gamma}, \mathbf{1}_{1 \times \gamma}]^{T},\nonumber\\
{\mathbf{z}}_{2} & = & [\mathbf{1}_{1 \times \gamma}, \mathbf{0}_{1 \times \gamma}]^{T},\nonumber\\
\label{eq:indf2}
\end{eqnarray}
\vspace{-.6cm}
\begin{eqnarray}
{\mathbf{z}}_{1} & = & [\mathbf{0}_{1 \times 2\chi}, \mathbf{1}_{1 \times \chi}, \mathbf{0}_{1 \times \chi}, \mathbf{1}_{1 \times \chi}, \mathbf{0}_{1 \times \chi}]^{T},\nonumber\\
{\mathbf{z}}_{2} & = & [\mathbf{1}_{1 \times \chi}, \mathbf{0}_{1 \times 4\chi}, \mathbf{1}_{1 \times \chi}]^{T},\nonumber\\
{\mathbf{z}}_{3} & = & [\mathbf{0}_{1 \times \chi}, \mathbf{1}_{1 \times \chi}, \mathbf{0}_{1 \times \chi}, \mathbf{1}_{1 \times \chi}, \mathbf{0}_{1 \times 2\chi}]^{T},\nonumber\\,
\label{eq:indf3}
\end{eqnarray} where $\gamma = \frac{N}{2}$ for the case of IRA, while for IRAA design $\gamma = \frac{N}{4}$. And, $\chi = \frac{N}{9}$ for the case of IRA, while for IRAA design $\chi = \frac{N}{15}$. In addition, for rate $R = \frac{1}{2}$ under IRAA design ${\mathbf{z}}_{i} = [{\mathbf{z}}_{i}, \mathbf{0}_{4 \times \gamma}]$, while for rate $R = \frac{1}{3}$ under IRAA design ${\mathbf{z}}_{i} = [{\mathbf{z}}_{i}, \mathbf{0}_{6 \times \chi}]$.

These indicator vectors are modelled on that of the original PEG
algorithm \cite{hu.05}, indicating sub-matrices for which placement
is permitted, thus imposing the form of (\ref{eq:pcm_ira_rc_r12}),
(\ref{eq:pcm_ira_rc_r13}), (\ref{eq:pcm_iraa_rc_r12}),
(\ref{eq:pcm_iraa_rc_r13}). The degree sequence as defined for LDPC
codes must be altered to take into account the structure imposed by
Root-Check codes, namely the identity matrices $\mathbf{I}$ and the
parity matrices $\mathbf{H}_{p}$, of (\ref{eq:pcm_ira_rc_r12}),
(\ref{eq:pcm_ira_rc_r13}), (\ref{eq:pcm_iraa_rc_r12}) and
(\ref{eq:pcm_iraa_rc_r13}). The pseudo-code for our proposed IRA-PEG
Root-Check algorithm is detailed in Algorithm \ref{alg:rcpegalg1},
where the indicator vector $\mathbf{z}_i$ is taken from
(\ref{eq:indf2}), (\ref{eq:indf3}) for constructing codes of rate
$R=\frac{1}{2}$, $R=\frac{1}{3}$ respectively.
\begin{algorithm}
 \caption{IRA-PEG Root-Check Algorithm}
 \label{alg:rcpegalg1}
\algsetup{
linenosize=\small,
linenodelimiter=.
}
\begin{algorithmic}[1]
\FOR{$j = 1 : K$}
\FOR{$k = 0 : D_s(j)-1$}
\IF{$j\ge\frac{N}{F}$ ~ \& ~ $k == 0$  }
\STATE Place edge at random among minimum weight CNs of the CN set permitted by the indicator ${\mathbf{z}}_{j}$.
\ELSE
\STATE Expand the PEG tree from the $\frac{(j-1)\cdot N}{F^2}$\emph{-th} VN to depth \emph{l} such that the tree contains all CNs allowed by the indicator vector \textbf{or} the number of nodes in the tree does not increase with an expansion to the \emph{(l+1)-th} level.
\STATE Place edge connecting the $\frac{(j-1)\cdot N}{F^2}$\emph{-th} VN to a CN chosen randomly from the set of minimum weight nodes which were added to the sub-tree at the last tree expansion.
\ENDIF
\ENDFOR
\ENDFOR
\end{algorithmic}
\end{algorithm}

\section{Simulations Results}

The performance of the proposed IRA and IRAA PEG Root-Check LDPC
codes when used in a Rayleigh block-fading channel with $F = 2$ and
$F = 3$ independent fading blocks is analysed. Moreover, we
considered the performance of such codes under fast fading channels.
All LDPC codes simulated have the same degree distribution under the
systematic  {parity} of PCM. Standard SPA algorithm is employed at
the decoder with a maximum of 20 iterations. Following
\cite{boutros.07, salehi.10}, a maximum of 20 iterations are enough
to obtain a good performance in terms of FER for fading channels.
The Gaussian outage limit in (\ref{eq:outage}) is drawn in dashed
line in each figure for reference. Our proposed IRA-PEG Root-Check
codes have a minimum girth of $12$.

\subsection{Performance for rate $R = \frac{1}{2}$}

\begin{figure}[htb]
 \centering
\resizebox{88mm}{!}{
\includegraphics{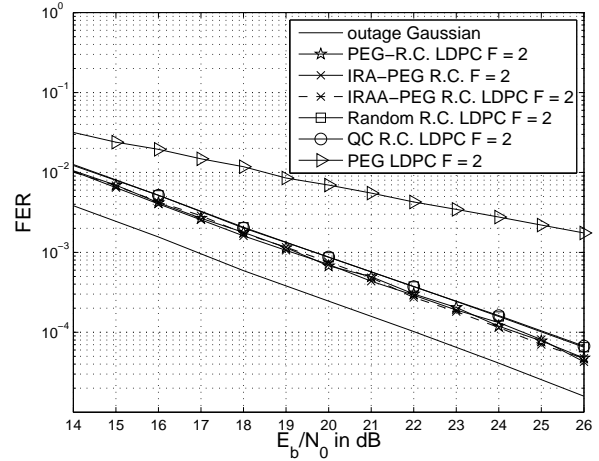}
} \vspace{-1.5em} \caption{FER performance for the IRA-PEG
Root-Check LDPC,  IRAA-PEG Root-Check LDPC, PEG Root-Check LDPC, QC
Root-Check LDPC, and PEG based LDPC codes over a block-fading
channel with $F = 2$ and $L = 1200$ bits. The maximum number of
iterations is 20.} \label{fig:fig3}
\end{figure}

In Fig. \ref{fig:fig3} it is compared the FER performance among the
proposed IRA-PEG Root-Check LDPC,  IRAA-PEG Root-Check LDPC, PEG
Root-Check LDPC from \cite{peg.comms.11}, QC Root-Check LDPC from
\cite{salehi.10} and PEG based LDPC \cite{hu.05} codes, all for
$R=\frac{1}{2}$. The codeword length is $L=1200$ bits. From the
results, it can be noted  that the proposed IRA(A)-PEG Root-Check
LDPC codes perform as  {well} as the PEG Root-Check LDPC code
design. The  {proposed} IRA(A)-PEG Root-Check codes bring the key
advantage of having less computational encoding complexity than the
other methods. Moreover, note that all Root-Check-based codes are
able to achieve the full diversity order of the channel, while the
non-root-check PEG LDPC codes fail to achieve full diversity.
 {ANDRÉ, WHAT ABOUT THE COMPLEXITY IN COMPARISON WITH
QC-BASED DESIGN? PLEASE COMMENT ON THAT.}

%

\subsection{Performance for rate $R = \frac{1}{3}$}

\begin{figure}[htb]
 \centering
\resizebox{88mm}{!}{
\includegraphics{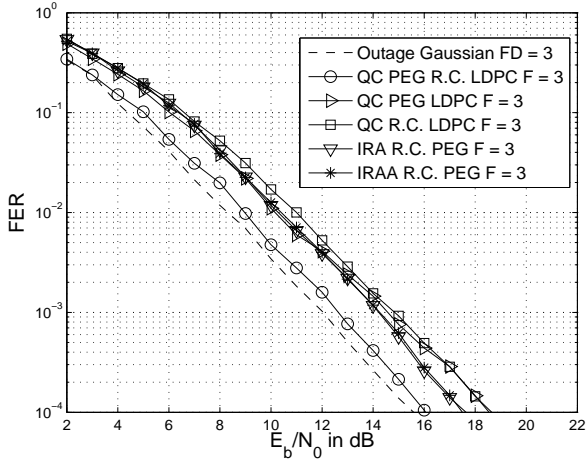}
} \vspace{-1.5em}\caption{FER performance for the IRA-PEG Root-Check
LDPC, IRAA-PEG Root-Check LDPC, QC-PEG Root-Check LDPC and QC-PEG
LDPC codes over a block-fading channel with $F = 3$ and $L = 900$
bits. The maximum number of iterations is 20.} \label{fig:fig5}
\end{figure}

In Fig. \ref{fig:fig5} it is compared the FER performance among the
proposed IRA-PEG Root-Check LDPC,  IRAA-PEG Root-Check LDPC, QC-PEG
Root-Check LDPC and QC-PEG LDPC, all for rate $R=\frac{1}{3}$. The
codeword length is $L=900$ bits.  Similar to the results for
$R=\frac{1}{2}$, it can be noted that the proposed IRA(A)-PEG
Root-Check LDPC codes perform as  {well} as the PEG Root-Check LDPC
code design.


%

\subsection{Performance for fast fading channels}

\begin{figure}[htb]
 \centering
\resizebox{88mm}{!}{
\includegraphics{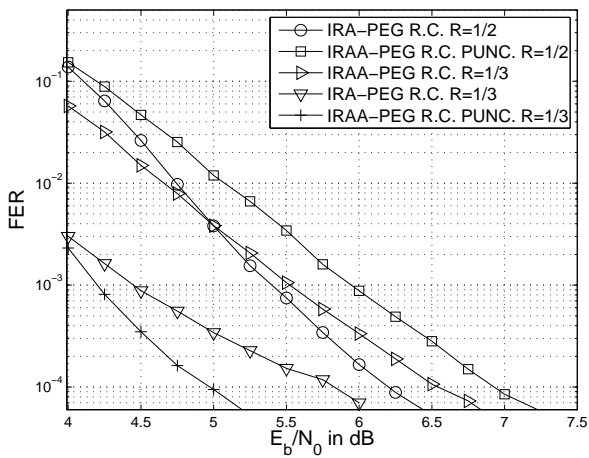}
} \vspace{-1.5em}\caption{FER performance for IRA-PEG Root-Check and
IRAA-PEG Root-Check codes with different code rates over a fast
fading channel.} \label{fig:fig7}
\end{figure}

In Fig. \ref{fig:fig7} it is compared the FER performance between
the proposed IRA-PEG Root-Check LDPC and IRAA-PEG Root-Check LDPC
codes for the case of fast fading and over different code rates.
From the legends in Fig. \ref{fig:fig7}, "PUNC." it means a
punctured version of IRAA design to obtain the same code rate as its
counterpart IRA code design. From the results, we can see that the
IRAA-PEG Root-Check design  {with rate $1/2$ is outperformed by the
IRA-PEG Root-Check design by about 0.75dB on average for the same
FER. Moreover, the non-punctured version of the IRAA-PEG Root-Check
design with rate $1/3$ was outperformed by the , the IRA-PEG
Root-Check design with rate $1/3$ by an average margin of 1.25dB.
Nevertheless, we can see that for the case of IRAA-PEG Root-Check
with rate $1/3$ punctured outperforms IRA-PEG Root-Check design with
rate $1/3$ by about 1dB for the same FER. In addition to the loss
observed by the IRAA-PEG Root-Check LDPC codes}, we have observed an
 { incredible feasibility - WHAT DO YOU MEAN BY THAT?
THIS EXPRESSION MUST BE REPHRASED. } to generate rate-compatible
codes.

\section{Conclusion}

A novel PEG-based algorithm has been proposed to design IRA(A)-PEG
Root-Check LDPC codes for $F = 2, 3$ fading blocks. Based on
simulations, the proposed method was compared to previous works in
\cite{iswcs.11, peg.comms.11, iswcs.12}. The results demonstrate
that the IRA(A)-PEG Root-Check LDPC codes generated by our proposed
algorithm perform as  {well} as the QC-PEG Root-Check LDPC codes
\cite{iswcs.12} in a wide range of SNR values and for fast fading
channel it can save up to 1dB. In addition, all designed Root-Check
LDPC codes presented in this paper are new in the essence that no
works were found with respect to IRA Root-Check design style.
Furthermore, we would like to reinforce that IRA encoding it is much
simpler than existing methods.  {Moreover, the proposed IRAA code
design brings flexibility in terms of rate and coding gains under
fast fading channels. There are some ongoing works in this are:
first, we are currently looking into the design of Root-Check codes
with accumulate repeat-accumulate (ARA) LDPC codes; second, we are
investigating the impact toward fast fading channels when the mother
code is a Root-Check LDPC code rate $1/2$ and by puncturing
techniques produce high rate codes; third, we are considering
improved decoding strategies \cite{vfap} for the above mentioned
designs.}

\section*{Acknowledgment}
This work was partially supported by CNPq (Brazil), under grant 237676/2012-5.

\bibliographystyle{IEEEtran}


\end{document}